# True random number generation through stochastic magnonic bistability

*Mengying Guo[1*], Zhenyu Zhou[1*], Denys Slobodianiuk[2], Roman Verba[2], Kristýna Davídková[3], Xueyu Guo[1], Xudong Jing[1], Yueqi Wang[1], Björn Heinz[4], Yiheng Rao[5], Carsten Dubs[6], Caihua Wan[7], Xiufeng Han[7], Andrii V. Chumak[3], Philipp Pirro[4], Qi Wang[1†]*

[1] *School of Physics, Hubei Key Laboratory of Gravitation and Quantum Physics, Institute for Quantum Science and Engineering, Huazhong University of Science and Technology, Wuhan, China*

[2] *V. G. Baryakhtar Institute of Magnetism of the NAS of Ukraine, Kyiv, Ukraine*

[3] *Faculty of Physics, University of Vienna, Vienna, Austria*

[4] *Fachbereich Physik and Landesforschungszentrum OPTIMAS, Technische Universität Kaiserslautern, Kaiserslautern, Germany*

[5] *Hubei Yangtze Laboratory, School of Integrated Circuits, Hubei University, Wuhan, China*

[6] *INNOVENT e.V., Technologieentwicklung, Jena, Germany*

[7] *Beijing National Laboratory for Condensed Matter Physics, Institute of Physics, Chinese Academy of Sciences, Beijing, China.*

True random number generators (TRNGs) underpin modern cryptography, yet existing implementations face fundamental trade-offs between speed, scalability, and entropy quality. Here, we demonstrate that stochastic switching in the bistable regime of spin-wave dynamics provides a physical entropy source for high-quality random number generation. Our magnonic random number generator (mRNG), based on a lithography-patterned microstrip on yttrium iron garnet (YIG), exploits thermal fluctuations near the nonlinear bistable regime to generate random bitstreams that pass all 15 NIST SP 800-22 statistical tests at rates with 20 Mb/s. We implement a random-bit multiplier using synchronized mRNG units and demonstrate scalability to 200-nm-wide nanoscale waveguides, establishing spin-wave bistability as a viable physical entropy source for integrated random number generation.

---

[*] These authors contributed equally to this work.
[†] Corresponding Author: williamqiwang@hust.edu.cn

**Introduction**

Reliable sources of true randomness are essential across science and technology, from cryptographic security to Monte Carlo simulations and probabilistic computing [1-3]. Conventional true random number generators (TRNGs) exploit thermal noise, oscillator jitter, or circuit metastability [4,5]. However, these approaches face trade-offs between speed, scalability, and entropy quality that limit their suitability for emerging applications in probabilistic computing and on-chip integration [6]. More advanced approaches - including quantum optical random number generators, spintronic devices based on magnetic tunnel junctions (MTJs), and memristor-based TRNGs - have been explored [1-3], yet they present their own challenges: quantum RNGs require expensive precision optics that are difficult to integrate on-chip, MTJ-based TRNGs demand complex multilayer fabrication and device patterning, and memristors suffer from limited endurance and poor device-to-device reproducibility. Crucially, most of these technologies require post-processing to pass standardized statistical tests for randomness due to bias or coherence in their original bitstreams.

These limitations highlight the need for alternative physical platforms that can intrinsically generate high-quality randomness while offering scalability, low power consumption, and compatibility with integrated architectures. In this context, wave-based and nonlinear dynamical systems provide a particularly attractive route, as they naturally exhibit complex, noise-sensitive behaviors that can be harnessed for entropy generation. Magnonics is an emerging field that utilizes spin waves (magnons) - collective excitations of electron spins in magnetic materials - to transmit and process information [7-10]. Unlike conventional electronics, magnonic systems offer unique advantages: ultra-low energy dissipation [11-13], nanoscale wavelengths [14-16], high-speed wave-based computing capabilities [17-19], controllable nonlinear phenomena [20-23], and compatibility with both quantum [24,25] and neuromorphic architectures [26-28]. These features position magnonics as a highly promising platform for next-generation computing and signal processing.

Bistability - a fundamental nonlinear phenomenon observed across diverse systems from optical resonators to biomolecular structures - describes a system's ability to maintain two distinct stable states under identical conditions, with controllable transitions between them [29-32]. Recent advances in magnonics have demonstrated robust bistability in nanoscale yttrium iron garnet (YIG) waveguides, revealing an exceptionally wide (>1 GHz) bistable window where spin-wave states can be deterministically switched via microwave pulses, enabling applications such as all-magnonic repeater and logic gates [33-35]. While previous studies have focused on deterministic switching, we uncover an intrinsic stochastic component in these transitions within specific excitation power regimes. This randomness arises from the interplay between the system's strong nonlinearity and thermal magnon fluctuations, where microscopic noise modulates the effective excitation threshold for spin-wave generation.

Building upon this phenomenon, we propose a compact and scalable magnonic random number generator (mRNG) with broad application potential. The device consists of a 5-μm-wide gold microstrip fabricated on an yttrium iron garnet (YIG) thin film using standard photolithography techniques. Its operation relies on the noise-sensitive excitation threshold of spin waves in the nonlinear region: identical microwave pulses applied to the microstrip generate stochastic spin-wave bursts due to threshold fluctuations driven by thermal noise in magnonic system. The raw bitstreams from the mRNG pass all 15 NIST SP 800-22 statistical tests for randomness. We further show that the concept is functional at the circuit level by implementing a random-bit multiplier from two synchronized mRNGs, and that it is scalable by fabricating waveguide devices down to 200 nm in width. Beyond random-number generation itself, the key distinction of this platform is that the random output can be emitted as a propagating magnon, opening a path towards directly embedding entropy into integrated magnonic hardware.

**Results**

Figure 1 schematically illustrates the parallel optical and electrical detection for characterizing magnonic random sequences. The core mRNG device consists of a 5-μm-wide gold microstrip antenna fabricated via photolithography on a 330-nm-thick YIG thin film [36]. An external magnetic field of 330 mT is applied out-of-plane z-axis to study forward volume spin wave propagation [37]. For spin wave excitation, the microstrip antenna is fed by two synchronized microwave signal generators combined through a power combiner, allowing to submit pump and trigger pulses of different frequency and duration for fine tuning of stochastic generation of propagating spin waves. The reflected microwave signal is routed sequentially through an amplifier, bandpass filter, variable attenuator, and fast diode before acquisition by the oscilloscope. A second 5-μm-wide gold microstrip, positioned 10 μm from the excitation antenna, detects the propagating spin waves generated by the mRNG. The detected signal then passes sequentially through an amplifier, bandpass filter, variable attenuator, and fast diode before being recorded on the same oscilloscope. Simultaneously, micro-focused Brillouin light scattering (μBLS) spectroscopy is used to measure the spin-wave intensity between the excitation and detection antennas [38].

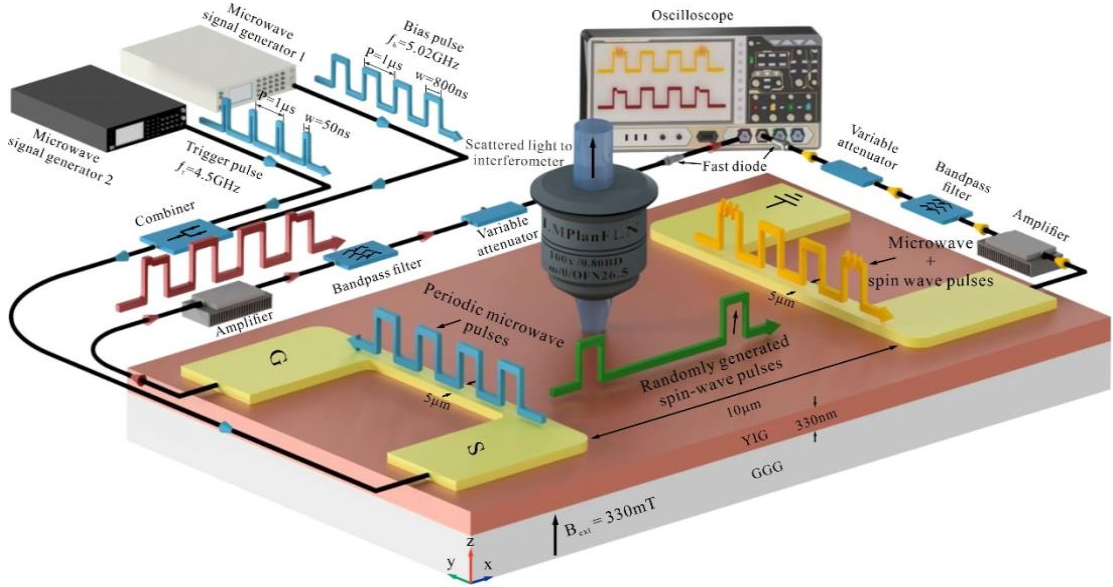

*Figure 1. **Parallel optical and electrical detection of magnonic random sequences.** The experimental setup consists of a 330 nm thick YIG layer as the magnon carrier, integrated with two 5 μm-wide gold antennas for spin wave excitation and detection. Two synchronized microwave generators - providing bias and trigger pulses - are combined and fed into one antenna to excite random magnonic pulses in the YIG. Reflected (red arrow) and transmitted (orange arrow) electrical signals are captured and analyzed using a real-time oscilloscope. Simultaneously, a micro-focused Brillouin light scattering (μBLS) optically probes the spin-wave intensity between two antennas. An external magnetic field (330 mT, applied along the z-axis) ensures the excitation of forward-volume spin-wave modes.*

To characterize the frequency range of the magnonic bistable window in this sample, a continuous microwave signal from Generator 1, sweeping the frequency between 4 to 5.8 GHz (in both directions) at a power of 36.3 mW, was applied to the microstrip antenna to excite spin waves. The focused laser spot of the μBLS was positioned midway between the two antennas to measure the spin-wave intensity. As shown in Fig. 2(a), the frequency sweep curves exhibit clear hysteresis, with the upward (black line) and downward (red line) failing to overlap, revealing a bistable frequency window of ~80 MHz similar to our previous studies [33-35]. Within this frequency range, the two distinct magnonic states coexist, exhibiting a 25-fold amplitude difference (approximately 14 dB).

We selected 5.02 GHz, located within the bistable window, as the bias pulse frequency (pulse width 800 ns, period 1 μs, power 36.3 mW). At this frequency, direct excitation of high-amplitude magnon state is not accessible when starting from a near-zero magnon population; the system remains in the low-amplitude state. To enable state switching, we apply additional trigger pulse at 4.5 GHz trigger pulse (outside the bistable window), consisting of 50 ns pulses with a 1 μs period and variable power. This trigger pulse deterministically excites spin waves at its own frequency, which, through nonlinear interactions, induce an upward shift of the spin wave spectrum and consequently of the bistable window.

As a result, the efficiently frequency gap is transiently bridged, allowing the bias pulse to efficiently inject energy into the magnonic system. This process leads to the rapid excitation of a high-amplitude state at the bias frequency - significantly stronger than the trigger-induced response - and drives the transition between the two stable states. Once established, the system remains in the high-amplitude state for the duration of the bias pulse [35].

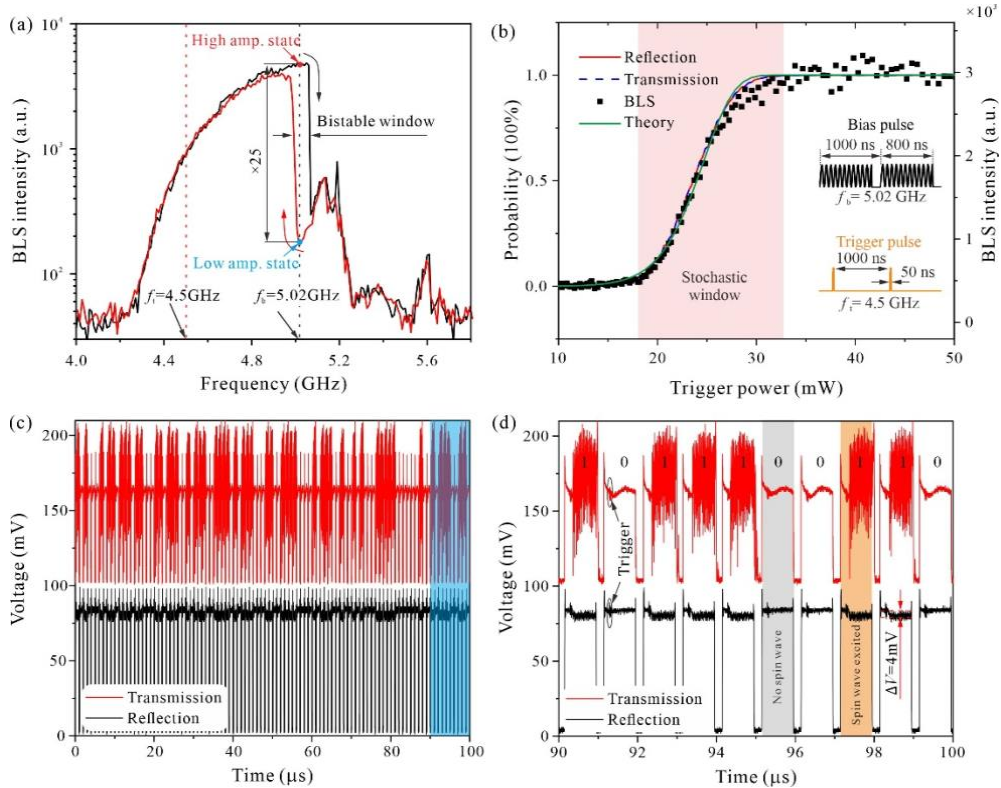

*Figure 2.* ***Stochastic behavior in magnonic bistable switching***. ***a***, *μBLS spectra for the up-frequency sweep (black line) and down-frequency sweep (red line) using continue microwave signal, revealing a magnonic bistable window with an 80 MHz frequency gap.* ***b****, Time-averaged μBLS intensity versus trigger power (black dots). The red solid line and blue dashed line indicate the probability of the magnonic random number sequence, extracted from electrical*

*transmission and reflection signals, respectively (see panel **c**). The green line shows the theoretical curve using Eq. (1). **c**, Transmission (red) and reflection (black) signals for the first 100 pulses. **d**, Zoomed-in view of the last ten pulse (highlighted in blue in **c**), displaying electrical transmission and reflection dynamics; "0" and "1" correspond to the absence/presence of switching event.*

Based on the above analysis, we initially expected to observe two distinct intensity levels in the BLS signal, separated by a sharp threshold as the trigger power increased. However, as shown by the black dots in Fig. 2(b) - which plots BLS intensity versus trigger power - the response displays a smooth sigmoidal transition rather than an abrupt jump. This deviation arises because each point in power was measured for ~30 seconds in the BLS experiment - corresponding to 30 million pulses - to ensure sufficient scattered photon statistics. As a result, the recorded BLS intensity represents an ensemble average over these pulses, and the smooth transition region between the two levels reflects the stochastic nature of the excitation process.

To validate our hypothesis, we simultaneously measured the electrical transmission and reflection signals for each individual pulse at a trigger power of 23.4 mW corresponding to the midpoint of the stochastic window with approximately 50% probability, as determined from an analysis of 10,000 pulses. Figure 2(c) presents the results of the first 100 pulses, demonstrating their stochastic behavior. Figure 2(d) shows a magnified view of the last ten pulses (highlighted by the blue region in Fig. 2(c)), where the transmission and reflection signals are plotted in red and black, respectively, to highlight finer details. As schematically illustrated in Fig. 1, all transmitted and reflected microwave pulses were processed through a fast diode, ensuring that only the envelopes of the signals were captured by the oscilloscope.

The gray region indicates one example of a pulse in which no spin wave excitation occurred. During this interval, the transmission signal arises primarily from microwave leakage between the two antennas [39], while the reflection signal remains nearly constant. A small peak observed around 100 ns delay (marked by the black arrows) corresponds to the trigger signal for each pulse. In the orange region, spin wave excitation is triggered, leading to a pronounced oscillation in the transmission signal. The observed low-frequency oscillations superimposed on the rectangular pulse originate from a beating effect between the directly coupled microwave signal and the propagating spin-wave signal. Due to strong nonlinear effects, the spin-wave frequency undergoes a slight amplitude-dependent shift during propagation, leading to a finite detuning from the excitation frequency. As a result, the interference between these two signals produces a beat note at the difference frequency, which is directly detected by the fast diode as a low-frequency oscillation. Simultaneously, the reflection signal exhibits a slight voltage drop due to energy transfer from the microwave to the spin wave. The average voltage during each pulse duration reveals two distinct levels (10 mV for transmitted signal and 4 mV for reflection signal; see the Supplementary Materials S1), enabling reliable and electrical discrimination between logic states "1" and "0" using threshold-based signal classification.

Figure 2(b) shows the spin wave excitation probability as a function of trigger power, with the red solid and blue dashed curves representing simultaneous measurements from transmission and reflection signals, respectively. Both datasets, extracted from 10,000 pulses at each power level, exhibit identical sigmoidal behavior and demonstrate excellent agreement with BLS measurements. These results confirm that magnonic bistable switching operates as a stochastic process within a well-defined power range (18 mW to 32 mW in this case). The physics underlying the stochastic switching arises from the interplay between the strong nonlinearity of the system and thermal fluctuations. In the absence of a trigger pulse and at zero temperature, the high-amplitude state is realized when the pump frequency $f_p$ lies below the left (low-frequency) edge of the bistability range $f_l$. The presence of thermal noise leads to fluctuations in the entire spin-wave spectrum, which in turn shifts the bistability edge according to $f_l \rightarrow f_l + \Delta f_n$, where $\Delta f_n$ is a random variable following a Gaussian distribution with zero mean and standard dispersion deviation $\sigma_f$, i.e., $p(\Delta f_n) \sim N(0, \sigma_f)$ (see Supplementary Materials S2 for justification and variance estimation). The application of a trigger pulse induces an additional frequency shift of the spin waves and consequently of the bistability range, due to nonlinear cross-mode frequency shift given by $\Delta f_{tr} = 2T_k |c_{tr}|^2$, where $T_k \approx \omega_M/2\pi$ is the nonlinear frequency shift coefficient and $c_{tr}$ is the amplitude of the trigger spin wave. The switching probability during the trigger pulse of the duration $t$ can then be expressed as

$$P(t) \approx 1 - \left(1 - \frac{1}{2}\mathrm{erfc}\left[\frac{f_{th} - 2T_k|c_{tr}|^2}{\sqrt{2}\sigma_f}\right]\right)^{tf_0}, \qquad (1)$$

where $f_{th} = f_p - f_l$ is the threshold value that must be overcome in the absence of a trigger pulse, $f_0$ is the FMR frequency and erfc is the complementary error function. In the absence of a trigger pulse, stochastic behavior is also possible; in this case $t$ corresponds to the duration of the pump pulse. Extracting from nonlinear FMR curve Fig. 2(a) the thermal variation of frequency $\sigma_f$=15 MHz and fitting the relation of trigger spin wave amplitude to the microwave power Power (mW) $\approx 5380|c_{tr}|^2$ (see details in Supplementary Materials S2), we get excellent agreement of the resulting theoretical curve (green solid line) with measurements, see Figure 2(b).

The stochastic switching window can be precisely tuned by adjusting either trigger duration $t$, the pump frequency $f_p$ or pump power, which affects the position of the bistability edge $f_l$, as it is clear form Eq. (1) (see extended experimental data in Supplementary Materials S3). This controllability is critical for cryptography applications, where unpredictability in both value and range strengthens security against attacks. This stochastic behavior is also consistently reproduced in micromagnetic simulations when thermal effects are incorporated (Supplementary Materials S4), further validating our experimental observations.

To evaluate the performance of our mRNG, we generated 85 million binary bits, segmented into 85 sequences, and subjected them to randomness evaluation using the NIST 800-22 statistical test suite [40]. The probabilistic bits (p-bits) successfully passed all 15 NIST tests, with full results provided in the Supplementary Materials S5. Therefore, the mRNG proposed here fulfills the criteria of a true random number generator (TRNG), leveraging intrinsic stochasticity in spin-wave dynamics. Notably, this is achieved on the raw bitstream without any post-processing or conditioning - a requirement that most hardware TRNGs, including those based on thermal noise, ring oscillator jitter, SRAM metastability, and memristive switching, are unable to meet [4-6].

In the above presented experimental data, a pulse with a period of 1 μs is used, enabling a p-bits rate of 1 Mb/s. Under the same experimental conditions, a shorter pulse period of 50 ns has been successfully tested, achieving a rate of 20 Mb/s (see Supplementary Materials S6). This rate can be further improved through material optimization (see Supplementary Materials S6), which enables operation at 1 Gb/s, as well as through equipment optimization or by employing parallel random bit generation (discussed later). To provide a quantitative performance reference for our device, we benchmarked it against state-of-the-art TRNG platforms, including CMOS, MTJ-based, and optical implementations (see Supplementary Materials S5 Table S2). The mRNG offers several fundamental advantages over competing hardware TRNG technologies. Unlike thermal noise and ring oscillator RNGs, which require extensive amplification chains and post-processing, the mRNG produces NIST-compliant random bits directly. Unlike quantum optical RNGs, which demand bulky and expensive single-photon detectors or homodyne setups, the mRNG relies on a simple lithographic microstrip on a YIG film. Compared to magnetic tunnel junction (MTJ) or memristor-based alternatives, the key advantage of the mRNG lies in its dual physical representation of randomness: the output p-bit can be encoded not only as an electrical voltage signal, but also directly as propagating magnons. These stochastic magnons can naturally propagate, interfere, and interact within magnonic circuits, enabling in-situ stochastic information processing. This capability opens a direct pathway toward hardware implementations of stochastic neural networks and probabilistic computing architectures, where randomness is not merely generated but intrinsically embedded into the computational medium. Moreover, the YIG-based platform provides effectively unlimited endurance, in contrast to memristors that degrade after $10^6$-$10^8$ switching cycles, and the excitation probability is continuously tunable - a degree of controllability not available in SRAM metastability or MTJ-based designs - it also generates propagating magnons in magnetic materials [1,2]. These magnons can interact with other magnons, enabling the realization of all-magnon integrated circuits that leverage magnon-generated random numbers for advanced computing applications [10,23].

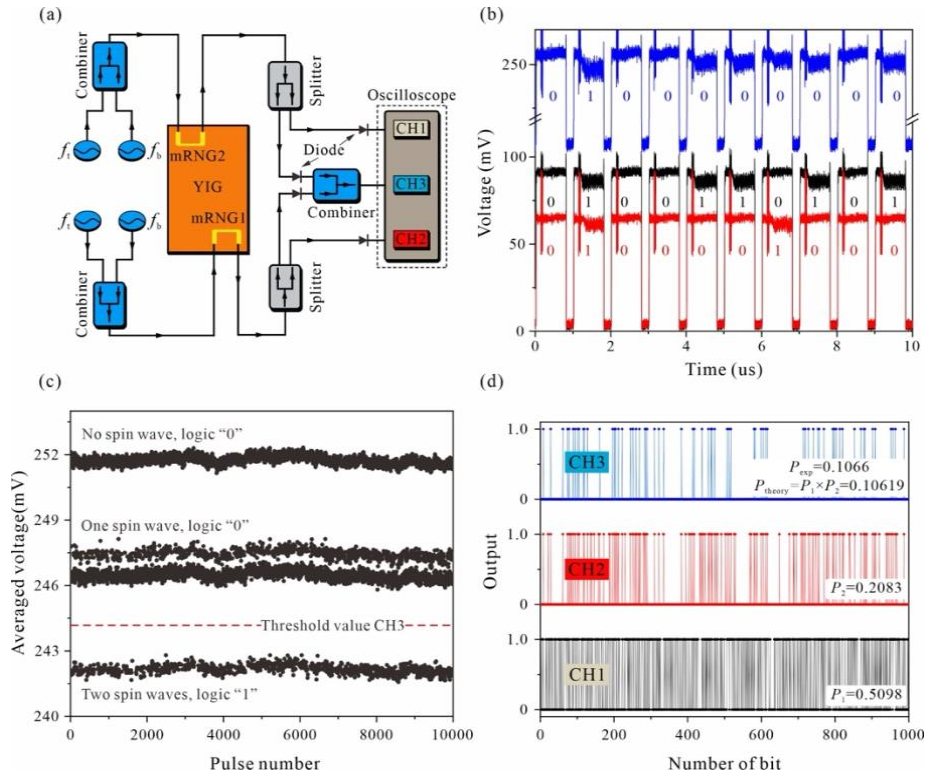

*Figure 3. **Working principle of magnonic random number multiplier**. **a**, Electrical schematic of the multiplier circuit integrating two mRNGs. **b**, Representative pulse sequences showing the reflection signals from RNG1 (black) and RNG2 (red), along with their combined output (blue). **c**, Statistical voltage distribution obtained from 10,000 pulses, highlighting three distinct operational states. **d**, 1,000-bit sequences extracted using the threshold-based signal classification across all measurement channels.*

Next, we discuss random number multiplication. Since transmission measurements requires a dual-antenna configuration, here we focused on the reflection signal (which gives the same results as is proven above); this approach simplifies the structure of the random number generator while maintaining measurement accuracy.

Figure 3(a) presents the electrical schematic of a magnonic random number multiplier integrating two mRNGs. Two independent 5-μm-wide gold antennas, fabricated on a YIG substrate using photolithography, function as mRNG elements that are spatially well separated and laterally offset, ensuring negligible coupling via both spin-wave propagation and microwave cross-talk. Each antenna is connected to two distinct microwave generators to apply trigger and bias pulses. The reflected signals from both mRNGs are divided into four paths: two signals are routed directly to oscilloscope channels CH1 and CH2 for independent measurement, while the remaining two are combined through a microwave combiner and acquired through CH3. This configuration enables simultaneous measurement of each mRNG's independent probability (via CH1 and CH2) and their joint probability product (via CH3). The probability characteristics of mRNG1 (CH1) and mRNG2 (CH2), measured concurrently, exhibit distinct sigmoidal responses to trigger level variations, consistent with the behavior observed in Fig. 2(b) (see Supplementary Materials S7).

To demonstrate multiplication of random probabilities, two distinct trigger power levels were selected, corresponding to probabilities of $P_1 \approx 0.5$ for mRNG1 and $P_2 \approx 0.2$ for mRNG2. Figure 3(b) shows ten pulse sequences from the three measurement channels. The red and black traces present the independent reflection signals from mRNG1 and mRNG2, respectively, while the blue trace corresponds their combined output. For CH1 and CH2, spin-wave excitation induces an immediate voltage drop in the reflection signal following the trigger, which we designate as logic "1", consistent with the behavior observed in Fig. 2(d).

Figure 3(c) presents the average voltage of CH3 over 10,000 pulse periods, revealing three well-defined voltage levels. The highest level (~252 mV) occurs when neither mRNG excites spin waves. Two intermediate levels, clustered around 247 mV, correspond to spin wave excitation in either mRNG1 or mRNG2, while the lowest level (~ 242 mV) indicates simultaneous excitation in both mRNGs. To perform the logic "AND" operation on the two stochastic inputs, we established a detection threshold ~244 mV for CH3. Signals below this threshold were classified as logic "1", while those above were assigned as logic "0".

Figure 3(d) displays the first 1,000-bit sample sequence across all three measurement channels, extracted from the full 10,000-pulse dataset using threshold-based signal classification. The independently measured probability of CH1 and CH2 were $P_1=0.5098$ and $P_2=0.2083$, respectively. The resulting product probabilities ($P_3=0.1066$) shows excellent agreement with theoretical value ($P_1 \times P_2=0.10619$). This result shows that stochastic magnons can be combined in hardware to perform probabilistic operations, rather than merely being read out as an external random sequence.

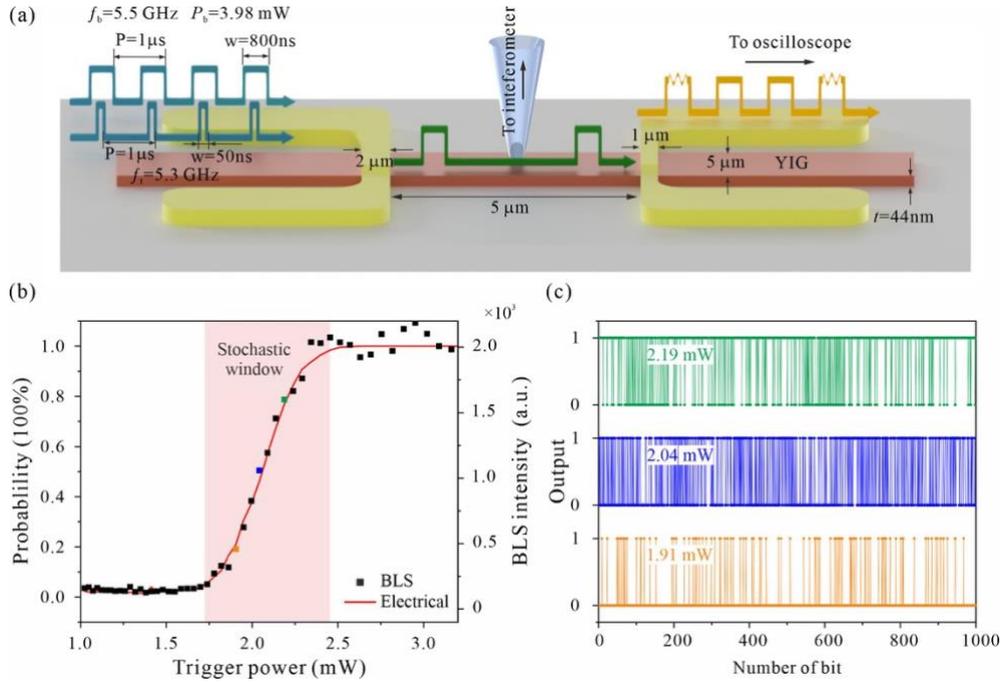

*Figure 4. **Scalability of the magnonic random number generator based on waveguides. a**, Schematic of the magnonic random bit generator implemented using magnonic waveguides with width of 5 µm. **b**, Electrical probability (red line) and BLS intensity (black dots) as a function of the trigger power for the waveguide based mRNG. **c**, Representative 1,000-bit sequences obtained at three different trigger powers, as indicated in **b**.*

To date, we have demonstrated magnonic random number generators based on YIG thin films and their application in stochastic multipliers circuits. To advance scalability and integration, we also investigate nanoscale waveguides as a promising alternative to thin-film architectures for miniaturized mRNG designs. Similar to Fig. 4(a), we fabricated waveguides with widths ranging from 200 nm to 20 µm (see Supplementary Materials S8 for fabrication details). The electrical connections retain the same configuration as in Fig. 1, with both electrical transmission signals and BLS used to characterize the spin wave intensity as a function of trigger power. For the 5 µm-wide waveguide, the bias and trigger pulses were set to 5.5 GHz (fixed power: 3.98 mW) and 5.3 GHz (varied power: 1-3.2 mW), respectively, with the same pulse width and period as before.

Figure 4(b) presents the electrical (red line) and BLS (black dots) signals measured from a 5 µm-wide waveguide, with a 5 µm separation between the excitation and detection antennas, demonstrating a characteristic sigmoid response that verifies stochastic behavior in a confined geometry. Figure 4(c) displays representative 1,000-bit sequences at three special power levels (marked in Fig. 4(b)), corresponding to probability of ~20%, ~50%, and ~80%, respectively. Similar behavior is observed in waveguides of other widths ranging from 200 nm to 20 µm, as detailed in the Supplementary Materials S8. Notably, a sigmoidal response in BLS measurements is observed even in the narrowest 200 nm waveguides, demonstrating the potential to scale mRNG devices down to the nanometer range. At the same time the electrical detection sensitivity is proportional to the magnetic volume under the antenna, restricting reliable spin-wave signal detection to waveguides with a minimum width of 5 µm using the current electrical method. This sensitivity limitation could potentially be improved by implementing more sensitive inverse spin Hall effect (ISHE) or magnetoresistance detection methods [13,41,42]. The nanoscale waveguide-based mRNG not only achieves significant miniaturization of individual devices but also enables parallel random bit generation by integrating multiple magnonic waveguides under a single input antenna [43,44], substantially increasing the p-bit generation rate.

One typical application scenario of p-bit is the physical encoding of the classic Ising machine algorithm, which is essentially a probabilistic annealing protocol. Nevertheless, the Ising machine can be encoded as a probabilistic bit (such as p-bit) or the phase of a wave (such as in Ring Oscillator). This mRNG system, stochastic generation of spin waves with both amplitude and phase degree of freedom, promisingly offers an ideal platform to achieve both modes as highlightened in Ref. [45] as an adoptive Ising machine.

**Discussion**

We have shown that stochastic switching of a nonlinear spin-wave system can be harnessed to build a compact magnonic random number generator. The device delivers raw bitstreams that pass the full NIST SP 800-22 suite, reaches 20 Mb/s in the present implementation, operates in circuit-level multiplier structures, and remains viable down to 200-nm-wide waveguides. The performance and scalability of this platform can be further enhanced for practical applications by simplifying the dual-pulse excitation system and introducing direct current to control the probability.

The present experiments use a trigger pulse to define a reproducible starting point for each switching event. That choice is convenient for controlled measurements, but it is not necessary to the effect. Stochastic switching can also be obtained with a single bias pulse whose onset-to-switching delay is itself random, provided that the readout is adapted accordingly (Supplementary Materials S9). This points to simpler future implementations in which the excitation scheme is reduced to a single pulse.

Additionally, the probability is adjusted by precisely controlling the microwave power. However, the requirement for precisely controllable microwave power sources also poses significant challenges for large-scale integration. Rather than modulating the microwave pulse power, an alternative approach involves fixing the microwave pulse power and introducing a direct current to adjust thermal noise via Joule heating (see the Supplementary Materials S10). A nanoscale, DC-tunable magnonic entropy source would combine compactness, in-situ tunability, and direct compatibility with propagating magnons, enabling their mutual interactions and paving the way for integrated all-magnonic devices based on magnon-generated random numbers. These results establish a compelling alternative to established TRNG technologies - combining the high entropy quality of quantum RNGs, the CMOS-compatible scalability of electronic TRNGs, and the wave-based computing integration underscoring the promise of magnonic systems in emerging computing architectures. Taken together, the mRNG achieves what no single existing technology provides: true stochastic randomness validated over 85 million bits without post-processing, a raw bit rate of 20 Mb/s that surpasses most hardware TRNGs by orders of magnitude, nanoscale dimensions (200 nm) rivalling state-of-the-art CMOS, and a unique dual output of both electrical signals and propagating magnons - particularly for stochastic computing, neuromorphic engineering, and hardware security, where robust and scalable physical entropy sources are essential.

**Data availability**

The data that support the plots presented in this paper are available from the corresponding authors upon reasonable request.

**Code availability**

The code used to analyze the data and the related simulation files are available from the corresponding author upon reasonable request.

**Acknowledgements**

This work was supported from the National Natural Science Foundation of China (Grant No. 12574118). R. V. and D. S. acknowledge support by the NAS of Ukraine via the project of KNU Department of Target Training No. 3F-2026. C. W. and X. H. acknowledge the Chinese Academy of Sciences President's International Fellowship Initiative (PIFI Grant No. 2025PG0006). P. P. acknowledges support by the Deutsche Forschungsgemeinschaft (DFG, German Research Foundation) -TRR 173–268565370 ("Spin + X", Project B01). C. D. acknowledges support by the DFG project under Grant No. 271741898. A. V. C. acknowledges the financial support of the Austrian Science Fund (FWF) by means of grant MagNeuro no. 10.55776/PIN1434524.

**Author Contributions**

M. G. fabricated the thin film sample, performed BLS measurements and micromagnetic simulations with help from X. G., X. J., and Y. W.. Z. Z. and M. G. performed electrical measurements. D. S. and R. V. provided explanation and developed theoretical model. K. D. and B. H. fabricate the nano/micro-scale sample. Y. R. and C. D. grew the YIG films. C. W. and X. H. contributed to the experimental design and discussions on the magnonic random number generation. A. C., P. P., and Q. W. led this project. Q. W. conceived the idea and Q. W., M. G. and Z. Z wrote the manuscript with the help of all the coauthors. All authors contributed to the scientific discussion and commented on the manuscript.

**Competing Interests**

The authors declare no competing interests.

**Correspondence** and requests for materials should be addressed to Q. W.